\documentclass[a4paper,french]{rnti}
\usepackage[ansinew]{inputenc}
\usepackage{graphicx}
\usepackage{array}
\usepackage{multirow}
\usepackage{url}

\titrecourt{Extraction d'entités dans des collections évolutives}

\titre{Extraction d'entités dans des collections évolutives}

\auteur{Thierry Despeyroux\affil{1},
        Eduardo Fraschini\affil{1}\\
        Anne-Marie Vercoustre\affil{1}}

\nomcourt{Th. Despeyroux et al.}

\affiliation{
INRIA - Rocquencourt\\
       Domaine de Voluceau\\
B.P. 105 - 78153 Le Chesnay Cedex\\
    \affil{1}
          prenom.nom@inria.fr\\
          \http{ http://www-rocq.inria.fr/who/Prenom.Nom/}\\
 }

\resume{%
Nous nous intéressons à l'extraction d'entités nommées avec
comme but d'exploiter un ensemble de rapports pour en extraire une
liste de partenaires. À partir d'une liste initiale, nous utilisons
un premier ensemble de documents pour identifier des schémas
de phrase qui sont ensuite validés par apprentissage supervisé sur
des documents annotés pour en mesurer l'efficacité avant d'être
utilisés sur l'ensemble des documents à explorer. Cette approche est
inspirée de celle utilisée pour l'extraction de données dans les
documents semi-structurés (wrappers) et ne nécessite pas de ressources
linguistiques particulières ni de larges collections de tests. Notre
collection de documents évoluant annuellement, nous espérons de plus
une amélioration de notre extraction dans le temps.
}

\summary{%
We are concerned by named entities extraction with the
final goal of constructing the list of partners found in an activity
report. Starting with an initial list of entities, we use a first set
of documents to identify syntactic patterns that are then validated in
a supervised learning phase on a set of annotated documents to perform
a performance test. The complete collection is then explored. This
approach comes from the one that is used in data extraction for
semi-structured documents (wrappers) and do not need any linguistic
ressources neither a large set for training. As our collection of
documents evoluate, we hope that the performance of the extraction
will become better year after year.
}

\begin{document}

\section{Introduction}

La reconnaissance et l'extraction d'entités nommées
cherche à localiser et à classer les
éléments atomiques d'un texte en catégories prédéfinies telles que
noms de personnes, organisations, localisation, dates, quantités,
valeurs monétaires, pourcentages etc.  Ce domaine de recherche est
très actif, bien que des outils commerciaux existent déjà.  
Citons, par exemple, REX\footnote{\url{http://www.basistech.com/entity-extraction/}}
(Rosette® Entity Extractor), Inxight SmartDiscovery\footnote{
\url{http://www.inxight.com/products/smartdiscovery/ee/}}, 
Convera-RetrievalWare Entity Extraction\footnote{\url{http://www.retrievalware.com/products/retrievalware/entity-extraction.asp}}
and Xerox-Research Entity Extraction
system\footnote{\url{http://www.ipvalue.com/technology/docs/Xerox_entity_extraction_pager.pdf}}.  
Le but de ces outils est d'annoter les documents avec des méta données (qui, quoi, où et quand) 
permettant une recherche d'information de plus haut niveau.

La plupart de ces systèmes demandent d'importantes ressources linguistiques 
(listes d'entités, larges corpus de référence) et l'écriture manuelle de règles pour s'adapter à un domaine particulier.
Par exemple, le système de Xerox utilise plus de 250 règles manuelles pour extraire des entités biologiques. 
D'autres règles sont inférées automatiquement et évaluées sur de gros corpus.

Du coté recherche, certains systèmes de reconnaissance d'entités
nommées utilisent des techniques à base de grammaires linguistiques, d'autres
des modèles statistiques. Les systèmes à base de
grammaires construits à la main obtiennent souvent de meilleurs résultats
au prix d'un travail très important par des linguistes chevronnés. 
Par ailleurs, les systèmes à base de modèles statistiques demandent
beaucoup de données d'apprentissage annotées, mais sont plus faciles à
porter vers d'autres langages, domaines ou genres de textes.




Nous proposons une approche dans laquelle, à partir d'une liste connue
d'entités, le système génère automatiquement des schémas de phrases
pouvant contenir ces entités.  Une étape d'apprentissages, à partir
d'un très petit nombre de documents permet de ne garder que les schémas
les plus pertinents. Cette approche s'inspire de celle utilisée pour
l'extraction de données dans des documents semi-structurés tels que
des pages Web (wrappers), basée sur la génération de programmes
d'extraction à partir d'un petit nombre d'exemples
(\cite{kushmerick00wrapper,276330,1135859,LermanJAIR,citeulike:494093}).  Au lieu de
s'appuyer sur les balises HTML des documents, nos règles s'appuient
sur les syntagmes du langage (balises linguistiques).  Cette approche
ne nécessite pas de ressources linguistiques particulières
(\cite{mcnamee02entity,cucerzan99language}) ni de larges collections de tests. 

Nous avons testé cette approche sur le rapport d'activité de l'Inria.
Il s'agit d'identifier, dans le rapport d'activité annuel, 
et plus particulièrement dans les sections décrivant 
les contrats de recherche et les relations internationales,
les organismes cités avec lesquels les équipes de
recherche coopèrent.
Dans ce contexte, identifier le plus possible de ces organismes ("rappel") est plus important qu'une précision élevée
puisque que la liste des entités extraite
peut être revue manuellement, même si cette tâche de vérification
s'avère très lourde pratiquement. D'autre part, ce genre de rapport
étant répétitif d'une année sur l'autre, et les partenaires évoluant lentement, il est intéressant que le
processus d'extraction puisse s'affiner avec le temps.

\section{Domaine applicatif}

Le rapport d'activité scientifique annuel de l'Inria est composé
d'environ 180 rapports en anglais
décrivant différents aspects de l'activité scientifique des équipe de recherche.
Depuis quelques années l'Inria est intéressée à exploiter cette source
riche d'information, disponible en XML.  Nous nous intéressons ici à
identifier les nombreux partenaires des équipes, en exploitant les
sections spécifiques décrivant les collaborations et les contrats.

Ce travail se heurte à plusieurs difficultés inhérentes à la
collection. Le style de ces sections est très peu homogène, parfois
télégraphique ou peu rédigé, avec une représentation des noms de
partenaires souvent approximative, voir avec des orthographes
erronées. Ces noms eux-mêmes peuvent être très divers: sigles plus ou
moins développés (FT R\&D), localisations intégrées au nom (Inria
Rocquencourt), noms d'organismes (EDF, MIT), de laboratoires
(LRI, LSR), de réseaux ou de noms de projets souvent confondus avec
des noms communs (Oasis, PARIS, Ondes). Le travail manuel d'annotation
des documents, utilisés pour la phase d'apprentissage et pour
l'évaluation, est une activité très coûteuses en temps et
intrinsèquement difficile.  Il serait exclu d'extraire les noms de ces
organismes à la main dans toute la collection (et chaque année).

Dans un premier temps, nous avons essayé d'utiliser un outil existant de bonne réputation, à savoir ANNIE, un des
composants du système GATE (\cite{ANNIE}) développé par l'université de
Sheffield (UK). Ce premier essai a été très décevant. Le taux de rappel était seulement de 0,23 si
nous cherchions la liste des organismes, et même de 0,17 si nous
cherchions toutes les occurrences des noms. Une des raisons est sans doute le style elliptique de cette partie du rapport, 
très différent du type de collections standards sur lesquels ANNIE est généralement validé (journaux, etc.). 
Nous avons donc décidé de développer une approche différente, partant des données réelles plutôt que de collections standards pour l'apprentissage.

\section{Méthode utilisée}

Les documents sont
au préalable annotés à l'aide de ANNIE qui détecte la fonction
grammaticale (nom, verbe etc.) des mots utilisés. Ce sont ces fonction
grammaticales (syntagmes) qui seront utilisés pour la construction de
schémas tel que nous en parlerons plus loin.

De façon standard, nous travaillons sur un ensemble réduit de rapports (collection test), 
utilisés dans la phase d'apprentissage et pour l'évaluation des résultats.
Dans ces rapports, les noms des organismes sont identifiés et annotés  à la main.
Cet ensemble est divisé en trois sous ensembles: 
le premier sous-ensemble, noté  $L$, sert à construire une liste initiale de noms
d'organisme\footnote{La liste de départ aurait pu être construite de façon arbitraire.};  le deuxième, $A$,  sert de base
d'apprentissage; le troisième, $B$, comme base de test. La table \ref{table:apprentissage} donne le nombre de documents et d'entités dans chaque sous-ensemble.

{\bf Les schémas de phrase (patterns)}

Tous les textes sont annotés (taggés) par les noms de syntagmes et, pour la collection test, 
par le tag <org> associés aux organismes identifiés. 
Par exemple, le fragment de phrase ``by Texas Instrument because'' devient:

{\small {\ttfamily <cat=''IN''>by <org>
<cat=''NNP''>Texas <cat=''NNPS''> Instruments </org>
<cat=''IN''>because}}

Les balises syntaxiques NNP, NNPS et IN indiquent respectivement un nom
propre, un nom pluriel, une préposition, et précèdent les
mots annotés. Nous avons ajouté la paire de balises {\tt <org>,
</org>} pour encadrer les noms d'organisme. Elles sont ajoutées manuellement
lors de la préparation des ensembles $L$, $A$ et $B$, et automatiquement
pendant l'exécution de l'algorithme d'apprentissage.

A partir de cette phrase, l'algorithme présenté plus loin va générer le schéma~:
 
\noindent {\ttfamily
IN$\sim$NNPS*\%$\sim$IN}. NNPS* est une nouvelle catégorie qui fusionne
NNP et  NNPS. Le signe ``{\ttfamily \%}'' indique une répétition
arbitraire du syntagme qui précède. Les caractères $\sim$ encadrent le nom d'entité à extraire.  Plus généralement, l'algorithme génère
des schémas pour toutes les occurrences d'organismes, avec un contexte gauche
et droit de longueur comprise entre 1 et 5. Ces schémas sont très
généraux puisque seule la catégorie syntaxique est retenue, et non sa
valeur (par ex. NNP et non pas University). Un schéma intuitivement intéressant est le schéma récursif 
{\ttfamily
'',''$\sim$NNPS*\%$\sim$'',''ORG}, qui, à partir d'un organisme connu dans une liste, va extraire les autres noms de la liste.

{\bf Algorithme d'apprentissage}

À partir de la liste d'organismes contenu dans
$L$, nous construisons, à partir des documents de $L+A$ une liste de schémas
$S_{LA}$ qui contiennent ces noms d'organisme. 

La phase d'apprentissage va déterminer les schémas les plus pertinents et supprimer ceux qui ne le sont pas.
Tout d'abord les schémas sont appliqués à $L+A$ pour extraire une liste d'organismes
potentiels $C$. Pour chaque schéma, nous calculons le nombre d'organismes de $C$ corrects ou incorrects,
et nous éliminons les schémas qui produisent trop peu de résultats corrects 
ou trop de résultats incorrects. 
Les schémas restants sont ensuite ordonnés en fonction de leur performance 
(nombre d'entités correctes/nombre d'entités incorrectes).

Ensuite, en partant de l'ensemble $A$ annoté avec les seules entités de la liste $L$,
nous appliquons un à un les schémas classés précédemment et nous extrayons de nouveaux organismes qui sont ajoutés à $L$ à l'itération suivante.
À chaque étape, la précision et le rappel sont évalués et 
l'algorithme s'arrête lorsque le rappel atteint un certain seuil, 
ou que la précision devient inférieure à un autre seuil.
En principe, le rappel va augmenter à chaque étape puisque les schémas les
plus performants sont ajoutés en premier. 
La précision, initialement égale à 1, puisque calculée 
à partir des seuls organismes de $L$ présents dans $A$, ne peut que se dégrader.


À cette étape du processus, nous avons donc sélectionné des
schémas d'extraction dont nous avons pu contrôler la performance sur
l'ensemble $A$. 

Dans la phase d'évaluation, nous appliquons les schémas précédemment sélectionnés
pour extraire les noms d'organisation de l'ensemble de test $B$ . Comme
$B$ a été lui aussi été annoté au préalable, nous pouvons
calculer la précision et le rappel pour valider notre approche.

\section{Expériences et résultats}

\begin{table}{\footnotesize
\begin{center}
\begin{tabular}{|l||l|l|l|l|}
\hline
& L & A & B & total\\
\hline\hline
Documents & 4 & 8 & 8 & 20\\
\hline
Entités différentes & 74 & 238 & 144 & 456\\
\hline
Occurrences d'entités & 238 & 418 & 271 & 827\\
\hline
\end{tabular}
\caption{\label{table:apprentissage}Nombre de documents et d'entités dans la collection initiale utilisée pour l'apprentissage et les tests (jeu de test 1)}
\end{center}}
\end{table}

Comme il est extrêmement fastidieux et difficile d'identifier les
organismes cités dans les documents, nous ne voulions pas avoir à le
faire pour plus de 20 documents. Afin de tester différents paramètres
de l'algorithme, nous avons effectué des permutations aléatoires des
documents dans les ensembles L, A et B afin de créer 10 jeux de test
différents.
Le tableau \ref{table:expériences} présente les résultats pour 3 de
ces jeux de test, pour des seuils d'apprentissage de 0,6 pour la
précision et  le rappel. Nous nous intéressons à la fois à
l'identification des noms d'organismes (comptage simple), et à
l'identification des occurrences de ces noms (comptage multiple).

\begin{table}
{\footnotesize
\begin{center}
\begin{tabular}{|m{3cm}||l|l|l|}
\hline
& Jeu de test 1 & Jeu de test 2 & Jeu de test 3 \\
\hline\hline
Entités dans L & 74 & 93 & 114\\
\hline
Entités de L trouvées dans A & 26 & 10 & 23\\
\hline

\multirow{3}{3cm}{Occurrences d'entités de L trouvées dans A} &
116 & 44 & 51\\
& (R=0,17; P=0,92) & (R=0,17; P=0,96) & (R=0,21; P=1)\\
& (MR=0,26; MP=0,94) & (MR=0,22; MP=0,89) &  (MR=0,25; MP=0,96)\\
\hline

Schémas retenus& 325 & 250 & 126\\
\hline

\multirow{3}{3cm}{Entités dans A à la fin de l'apprentissage} &
247 & 183 & 126\\
& (R=0,56; P=0,60) & (R=0,56; P=0,60) & (R=0,61; P=0,90)\\
& (MR=0,67; MP=0,63) & (MR=0,67; MP=0,62) &  (MR=0,64; MP=0,91)\\
\hline

Entités de L trouvées dans B & 12 & 24 & 17\\
\hline

Entités extraites de B & 140 & 330 & 240\\
\hline

Comptage simple & R=0,31; P=0,46 & R=0,45 ;P=0,41 & R=0,36; P=0,47\\
\hline

Comptage multiple & R=0,35; P=0,44 & R=0,53 ;P=0,41 & R=0,42; P=0,49\\
\hline

\end{tabular}
\caption{\label{table:expériences}Résultats pour 3 jeux de test; R, P: rappel et précision (comptage simple) MR, MP: comptage multiple (occurrences)}
\end{center}
}
\end{table}

On peut tout d'abord remarquer que le rappel de départ pour l'ensemble
$A$ est faible (0,17 et 0,21), ce qui indique une grande diversité de
partenaires selon les différentes équipes. Bien que la précision de
départ devrait être égale à 1 nous voyons que ce n'est pas tout à fait
le cas. En effet par absence de normalisation si $L$ contient ``FT''
et un document de $A$ ou $B$ contient ``FT R\&D'' identifié comme nom
d'organisme, FT sera identifié mais considéré comme non valide pour la
calcul de la précision et du rappel.

Le rappel à la fin de la
période d'apprentissage a été multiplié par plus de 2 en moyenne pour
le comptage simple, mais reste malgré tout assez faible. Il faut
rappeler que l'algorithme d'apprentissage s'arrête lorsque le rappel
est plus grand que 0,6 ou la précision inférieure à 0,6 (pour le
comptage multiple). On ne peut donc pas espérer des
valeurs très élevées à la fin de l'apprentissage, en utilisant des
schémas génériques et calculés automatiquement.

Nous avons évalués les résultats sur 5 jeux de test identiques pour
3 couples de seuils différents. Le tableau \ref{table:seuils} montre
les résultats moyens sur les 5 jeux de test.

On peut voir que les seuils ont une influence non seulement sur la
précision finale, mais aussi sur le nombre de schémas validés par
apprentissage. Ce nombre est plus petit si la précision demandée est
élevée, ce qui avantage le temps d'extraction. En contrepartie le
nombre d'entités extraites est inférieur ce qui est inconvénient pour
notre application.

Il se trouve que la partie ``contrats'' contient plus facilement des
partenaires industriels et la partie ``collaboration'' plus souvent
des partenaires académiques.  Nous avons donc faits des expériences 
en effectuant un apprentissage séparément sur chacun de ces groupes.
Les résultats sont meilleurs pour la partie
``collaborations'', les noms d'universités étant plus facile à
identifier, mais contrairement à notre attente, les résultats sont moins
bons quand on traite les parties ``contrats'' et ``collaborations'' de
façon séparée plutôt qu'ensemble.

Finalement, nous avons appliqué les schémas sélectionnés à l'ensemble des
180 rapports. 
Selon les expérimentations (non reportées ici, faute de place), 
1500 à 3000 noms ont été extraits. Un essai de validation d'une liste de
1500 noms a montré la difficulté d'une telle tâche.

\begin{table}{\footnotesize
\begin{center}
\begin{tabular}{|l||l|l|l|l|l|}
\hline
Seuils& R mult & P mult & R simple & P simple & Nb de schémas\\
\hline\hline
R=0,4; P=0,7 & 0,30 & 0,62 & 0,24 & 0,62 & 59\\
\hline
R=0,6; P=0,6 & 0,44 & 0,48 & 0,37 & 0,48 & 228\\
\hline
R=0,7; P=0,4& 0,48 & 0,44 & 0,41 & 0,43 & 250\\
\hline
\end{tabular}
\caption{\label{table:seuils}Précision et rappel final suivant le choix des seuils}
\end{center}}
\end{table}

\section{Conclusion}

Nous avons présenté une méthode pour extraire les noms d'organismes dans
des parties 
de documents assez peu rédigées.
Notre approche s'inspire des méthodes inductives des extracteurs pour des
documents semi-structurés, et ne requière pas d'importantes ressources
linguistiques ni de mise au point manuelle.
Les résultats, bien qu'un peu décevants, montrent qu'il est possible de
découvrir un grand nombre d'organismes non connus à l'avance.

D'une année sur l'autre, il y a une certaine continuité dans les
partenaires avec lesquels les équipes Inria travaillent. Il est donc
raisonnable d'utiliser la liste des organisations d'une année pour
initialiser l'extraction d'entités pour l'année N+1. 
Même si les listes produites demandent à être validées manuellement, c'est
certainement plus rapide que d'extraire manuellement le nom
des organismes à partir des 180 rapports d'activité. 
 
Nous envisageons également de générer, en plus de schémas purement
syntaxiques, des schémas
plus spécifiques tirant partie des valeurs de certains syntagmes.

\bibliographystyle{rnti}

\bibliography{etam}

\providecommand\Fr{}
\providecommand\Eng{}
\providecommand\andname{and}
\providecommand\andnamec{and}

\begin{thebibliography}{}


\bibitem[{Adelberg}(1998){Adelberg}]{276330}
Adelberg, B. (1998).
\newblock Nodose - a tool for semi-automatically extracting structured and
  semistructured data from text documents.
\newblock In {\em SIGMOD '98, ACM SIGMOD international conference on Management
  of data}, New York, NY, USA, pp.\  283--294. ACM Press.

\bibitem[{Cucerzan \andnamec{} Yarowsky}(1999){Cucerzan \andnamec{}
  Yarowsky}]{cucerzan99language}
Cucerzan, S. \andname{} D.~Yarowsky (1999).
\newblock Language independent named entity recognition combining morphological
  and contextual evidence.
\newblock In {\em 1999 Joint SIGDAT Conference on EMNLP and VLC}.

\bibitem[{Cunningham et~al.}(2002){Cunningham, Maynard, Bontcheva, \andnamec{}
  Tablan}]{ANNIE}
Cunningham, H., D.~Maynard, K.~Bontcheva, \andname{} V.~Tablan (2002).
\newblock Gate: A framework and graphical development environment for robust
  nlp tools and applications.
\newblock In {\em 40th Anniversary Meeting of the Association for Computational
  Linguistics (ACL'02)}.

\bibitem[{Irmak \andnamec{} Suel}(2006){Irmak \andnamec{} Suel}]{1135859}
Irmak, U. \andname{} T.~Suel (2006).
\newblock Interactive wrapper generation with minimal user effort.
\newblock In {\em WWW '06, 15th international conference on World Wide Web},
  New York, NY, USA. ACM Press.

\bibitem[{Kushmerick}(2000){Kushmerick}]{kushmerick00wrapper}
Kushmerick, N. (2000).
\newblock Wrapper induction: Efficiency and expressiveness.
\newblock {\em Artificial Intelligence\/}~{\em 118\/}(1-2), 15--68.

\bibitem[{Lerman et~al.}(2003){Lerman, Minton, \andnamec{}
  Knoblock}]{LermanJAIR}
Lerman, K., S.~Minton, \andname{} C.~Knoblock (2003).
\newblock Wrapper maintenance: A machine learning approach.
\newblock {\em Journal of Artificial Intelligence Research\/}~{\em 18},
  149--181.

\bibitem[{Liu et~al.}(2003){Liu, Grossman, \andnamec{} Zhai}]{citeulike:494093}
Liu, B., R.~Grossman, \andname{} Y.~Zhai (2003).
\newblock Mining data records in web pages.
\newblock In {\em KDD '03: Proceedings of the ninth ACM SIGKDD international
  conference on Knowledge discovery and data mining}, New York, NY, USA, pp.\
  601--606. ACM Press.

\bibitem[{McNamee \andnamec{} Mayfield}(2002){McNamee \andnamec{}
  Mayfield}]{mcnamee02entity}
McNamee, P. \andname{} J.~Mayfield (2002).
\newblock Entity extraction without language-specific resources.
\newblock In {\em CoNLL-2002}.

\end{thebibliography}

\end{document}